# Effect of Ambient Gas Pressure on the Thermal Exfoliation of Graphite Oxide: Tuning the Number of Graphene Sheets

Himanshu Raghubanshi, M. Sterlin Leo Hudson and O. N. Srivastava*

## Abstract

We have investigated the effect of ambient gas pressure (Argon) used during thermal exfoliation of graphite oxide on the number of graphene sheets in the resulting graphene. The ambience gas Ar pressures used are 80, 25, 1 atmosphere and $10^{-3}$ as well as $10^{-5}$ torr. We have found curious result that the number of graphene sheets in exfoliated graphene stack is dependent on the applied pressure for exfoliation of graphite oxide (GO). By lowering the exfoliation pressure, the numbers of sheets get reduced. The numbers of sheets in exfoliated graphene stack have been determined by Lorentzian fit of their respective 002 peak of XRD profile. The numbers of sheets have also been confirmed with high resolution transmission electron microscopy (HRTEM). The numbers of sheets in exfoliated graphene stack samples have been found to be 28, 12, 8, 5 and 3 at 80 atm, 25 atm, 1 atm, $10^{-3}$ torr and $10^{-5}$ torr pressures respectively. The results of present investigation suggest a simple but effective method for synthesizing graphene with a specific number of sheets.

## 1. Introduction

Graphene is the single layer hexagonal pattern of $sp^2$ bonded carbon atoms that are densely packed in honeycomb crystal lattice. The first report on the synthesis of 2D network of carbon atoms such as graphene has appeared in 2004 through the pioneering work of Novoselov et al. 2004. They have prepared graphene by the removal of graphitic sheets layer by layer from ordered pyrolytic graphite employing "scotch tape". Subsequently, several methods have been developed for the preparation of graphene such as pyrolysis of camphor under reducing conditions (Somani et al. 2006), exfoliation of graphite oxide (McAllister et al. 2007), solvothermal synthesis (Choucair et al. 2009).

Recently, graphene has received considerable attention in the scientific and industrial community due to its extraordinary properties which lead to several potential applications (Geim et al. 2007). Ballistic electron transport, linear current-voltage (I-V) characteristic, huge sustainable currents ($>10^8$ A/cm$^2$), anomalous quantum Hall effect at room temperature and fractional quantum Hall effect at low temperatures and superior mechanical properties of graphene makes them as a high performance engineering material for several applications (Novoselov et al. 2004 and Wilson et al. 2006). The significant applications of graphene are as nanofillers in nanocomposites, sensors, transparent electrodes, superior electronic devices, chemical detectors, electrochemical capacitors and energy storage etc. (Stankovich et al. 2006, Ryu et al. 2008 and Verdejo et al. 2008).

The method which can give large yield of graphene in a single run corresponds to thermal exfoliation of graphite oxide. Natural graphite is generally used as a starting material where the forces holding the parallel graphene sheets are weak Van der Waals forces. The intermediate compound formed in these methods is graphite oxide (GO), an oxygen rich carbonaceous layered material, dispersible in water. GO has often been made by acid treatment of graphite powder via three principal methods such as Hummers method (Hummers et al. 1958), Brodie (Brodie 1859), and Staudenmaier (Staudenmaier 1898). Each layer of GO is essentially an oxidized graphene sheet commonly referred to as graphene oxide. Complete oxidation of graphite for the formation of GO is an essential condition for the synthesis of bulk quantities of graphene sheets (Schniepp et al. 2006). The GO is suddenly heated by inserting it in a pre-heated furnace at 1050 °C. Under this type of heat treatment GO gets exfoliated into graphene sheets. Graphene prepared by this method invariably contain some oxygen functionalities. These are mainly hydroxyl and epoxy functional groups. In view of this graphene prepared from thermal exfoliation of graphite oxide is often termed as functionalized graphene sheet (FGS). For reduction of the graphite oxide (GO), thermal treatment is the most suitable one, since it provides high yield of graphene sheets. The graphene sheets obtained by this method are functionalized graphene sheets. Some oxygen functionalities are still attached to the graphene sheets. The mechanism of exfoliation is mainly the production and expansion of $CO_2$ evolved between the graphene sheets during rapid heating (Schniepp et al. 2006). This allows gases to diffuse out of the galleries instead of heating up and creating sufficient pressure to overcome van der Waals interactions between graphene sheets. So far graphene sheet from GO have been obtained through exfoliation under atmospheric inert gas e.g Ar pressure. Studies of graphene are not limited to one-atom thick single-layer graphene alone but also include bi- and few-layer (<10 sheets) graphenes (Rao et al. 2010).

In the present investigation, exfoliation of GO has been carried out under variable pressures such as at high pressure (80 and 25 atm Ar gas), atmospheric pressure (1 atm Ar gas) and low pressure ($10^{-3}$ and $10^{-5}$ torr pressure). Interestingly, we found that the formation of number of sheets in the exfoliated graphene is dependent on the applied exfoliation Ar pressure. Thus the number of sheets of graphitic sheets in exfoliated graphite under Ar gas pressure of 80, 25, 1 atmosphere and $10^{-3}$ as well as $10^{-5}$ torr are 28, 12, 8, 5 and 3 respectively. To the best of our knowledge, this is the first investigation of its type where numbers of graphene sheets have been tuned by the variation of ambient gas pressure during thermal exfoliation of graphene. Thus the method described here may form an effective process for synthesizing graphene with a pre-determined number of sheets.

## 2. Experimental Techniques

### 2.1. Exfoliation Method

The GO was prepared by the Staudenmaier 1898 method. Pure graphite (Sigma Aldrich, 99.99%) was oxidized through acid treatment of conc. $H_2SO_4$, conc. $HNO_3$ and $KClO_3$. For thermal exfoliation at high pressure, the dried GO (0.2 g) is charged into a steel reactor which is filled with Ar at different pressures. This system was subjected to rapid heating to 1050 °C for 3 minute with continuous water flow into the reactor for cooling purpose. The as-obtained material was exfoliated graphite oxide (EGO) or functionalized graphene sheets (FGSs) at 80 and 25 atm Ar pressures. For thermal exfoliation at atmospheric pressure in Ar ambience, the dried GO (0.2 g) is charged into a long quartz tube (1.5 m) with diameter 2.5 cm and purged with argon. Rapid heating to 1050 °C for 30 sec gives FGSs. For thermal exfoliation at low pressure, the dried GO (0.2 gm) is charged into a steel reactor which was evacuated up to $10^{-3}$ torr by using rotary and diffusion pump. Again this system was subjected to rapid heating at 1050 °C for 3 minute with continuous water flow into the reactor for cooling purpose. The as-obtained material corresponds to FGSs at $10^{-3}$ torr pressure. Similar process was made for $10^{-5}$ torr Ar pressure.

### 2.2. Characterization Techniques

The structural characterization of the samples were carried out using X-ray diffraction technique employing X'Pert PRO PANalytical diffractometer equipped with graphite monochromator with a Cu source ($\lambda$=1.54 Å, CuK$\alpha$ operating at 45 KV and 40 mA). HRTEM was performed using TECNAI G$^2$, operating at 200 KV in diffraction & imaging modes. The samples for TEM were prepared by ultrasonically dispersing the powder in a mixture of ethanol and deionized water and then coating the above dispersing solution onto copper grids.

## 3. Results and Discussion

### 3.1. XRD Analysis

XRD patterns of as-synthesized GO, and graphene sheets at different conditions are presented in Fig. 1. Fig. 1A shows the XRD pattern of GO, the sharp nature of GO peak (00.2) suggests that it is crystalline in nature. Fig. 1(B-F) shows the shift in d values from GO to EGO, which indicates the successful exfoliation (thermal reduction) for all ambient gas pressure employed for exfoliation in the present study. From Fig. 1, it is clear that when we move towards lower exfoliation pressure, the broadening of (00.2) peaks increases. The higher the broadening of (00.2) peak, the lower the number of graphene sheets. For single graphene sheet all the diffraction peak will be eliminated (Verdejo et al. 2008). The comparative sharp graphitic reflection of the 002 peak in the XRD pattern of Fig. 1B (exfoliation pressure 80 atmosphere Ar) shows that it consists of comparatively number of sheets.

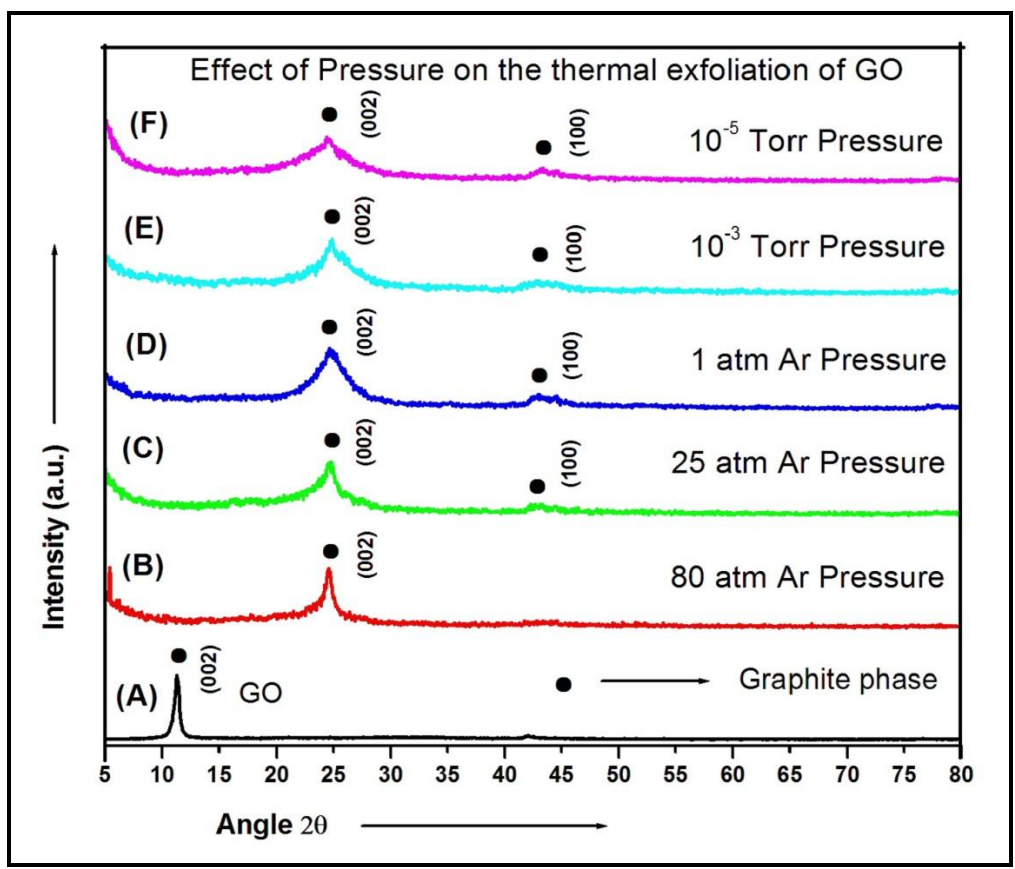

**Fig. 1:** Typical XRD pattern of (A) as-synthesized GO, (B-D) graphene sheets at high Ar gas pressure 80, 25 and 1 atm respectively, (E & F) graphene sheets at low pressure ($10^{-3}$ and $10^{-5}$ torr pressure respectively).

Fig. 1(C-E) shows relatively broad peak and Fig. 1F shows broadest peaks than the other peaks presented in the XRD pattern (Fig. 1).

## 3.2. Lorentzian Fit

We have fitted the (00.2) profiles to obtain the number of graphene sheets in the EGO samples obtained for exfoliation carried out under different Ar gas pressures. By doing Lorentzian fitting of the (00.2) reflection, one can obtain the average number of sheets using the Debye-Scherrer formula. Fig. 2 shows representative typical Lorentzian fit for the (00.2) reflection. In this figure we have shown the Lorentzian fit for the case of exfoliation under 80 and 1 atmosphere pressure of argon and $10^{-3}$ torr. Similar Lorentizian fits have been obtained for the other graphene stacks obtained under 25 atm and $10^{-5}$ torr pressure.

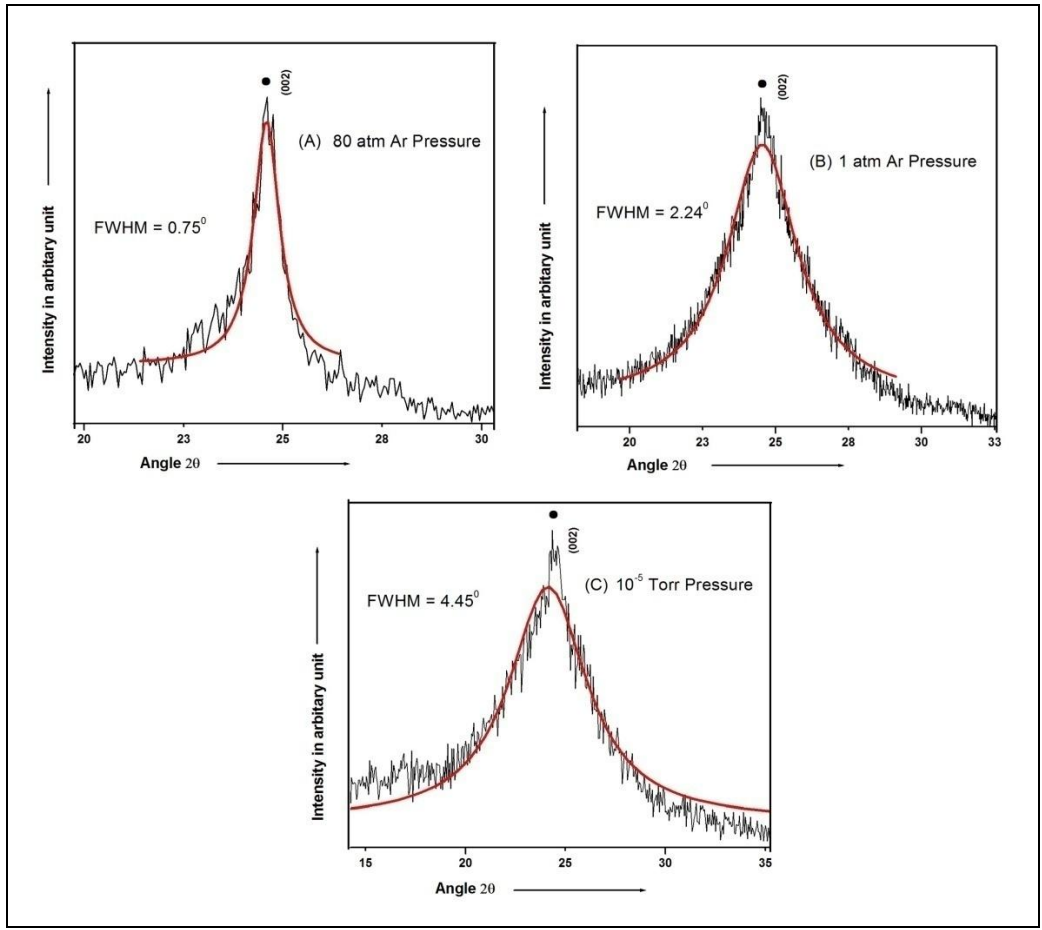

**Fig. 2: Lorentzian Fit for the (00.2) diffraction peaks for (A) 80 atm Ar pressure, (B) 1 atm Ar pressure and (C) $10^{-5}$ torr pressure.**

From Debye-Scherrer formula, we have calculated the average number of sheets for EGO samples obtained at 80, 25 and 1 atm and $10^{-3}$ and $10^{-5}$ torr pressure. The calculated number of sheets in the exfoliated graphene stack is 28, 11, 8, 5 and 4 sheets.

The numbers of the sheets in EGO were calculated by using the classical Debye-Scherrer equations: Thickness (t) of the graphitic stack is given by

$$t = 0.89\lambda/\beta_{002}\cos\theta_{002}$$

The number of Sheets is given by

$$N = t/d_{00.2}$$

Where t is the thickness of graphitic stack, β is the full width at half maximum (FWHM) and it was obtained from the Lorentzian fit for the (00.2) profile, N is the number of graphene sheets, $d_{002}$ interlayer spacing between the (00.2) planes.

Table 1 shows relevant parameters and the number of sheets calculated (XRD) and observed from HRTEM analysis. The number of sheets calculated from the two methods (Lorentzian fit of XRD profile and HRTEM analysis) exhibit a good match.

**Table 1**

| S.No. | Exfoliation of GO at Pressure | Angle $2\theta_{002}$ (deg) | Interlayer spacing $d_{002}$ (A°) | FWHM (Lorentzian Fit) (deg) | $\beta_{002}$ (radian) | Thickness (T) A° | Number of Sheets N (Calculated) From XRD | Number of Sheets N (Observed) HRTEM |
|---|---|---|---|---|---|---|---|---|
| 1. | 80 atm Ar | 24.54 | 3.62 | 0.75 | 0.013 | 108.60 | ~30 | 28 |
| 2. | 25 atm Ar | 24.61 | 3.62 | 1.61 | 0.028 | 50.68 | ~14 | 12 |
| 3. | 1 atm Ar | 24.57 | 3.62 | 2.24 | 0.039 | 36.20 | ~10 | 8 |
| 4. | $10^{-3}$ torr | 24.59 | 3.62 | 3.78 | 0.066 | 21.72 | ~6 | 5 |
| 5. | $10^{-5}$ torr | 24.44 | 3.64 | 4.45 | 0.078 | 18.20 | ~5 | 3 |

Here it should be mention that the number of sheets calculated from Debye-Scherrer formula gives the approximate result because it depends on the width of the XRD profile. However HRTEM gives the accurate results because it gives the direct evidence of the number of sheets presented in the sample.

### 3.3. HRTEM Analysis

In order to further check on the number of sheets in the exfoliated graphene stack high resolution electron microscopy (Fig. 3) was carried out. Fig. 3A shows the HRTEM of graphene sheets obtained at 80 atm Ar pressure. It reveals that the stacked graphene contain 28 graphene sheets. Inset of Fig. 3A shows its magnified version. Fig. 3B shows the graphene sheets obtained at 25 atm Ar pressure, it reveals that the stacked graphene contain 12 graphene sheets. Inset of Fig. 3B shows its magnified version. From the Scherrer formula, we have obtained 30 and 14 graphene sheets for 80 and 25 atm Ar pressure respectively, which matches well with the number of graphene sheets obtained from HRTEM.

Fig. 3C shows the HRTEM of graphene stacked obtained at 1 atm Ar pressure (atmospheric pressure), it reveals that the stacked graphene contain 8 graphene sheets (sheets). Inset of Fig. 3C shows its magnified version. From XRD profile fitting also we have got the nearly same number of graphene sheets i.e. 10.

Fig. 3(D&E) shows the HRTEM of graphene stacked sample obtained at $10^{-3}$ and $10^{-5}$ torr pressure. This reveals that the stacked graphene contains 5 and 3 graphene sheets, respectively. Inset of Fig. 3A shows its magnified version. The number of graphene sheets calculated from the XRD profile fitting for $10^{-3}$ and $10^{-5}$ torr exfoliation pressure is 6 and 5 respectively. Thus there is a very good match between the HRTEM and XRD result in regards to the number of graphene sheets.

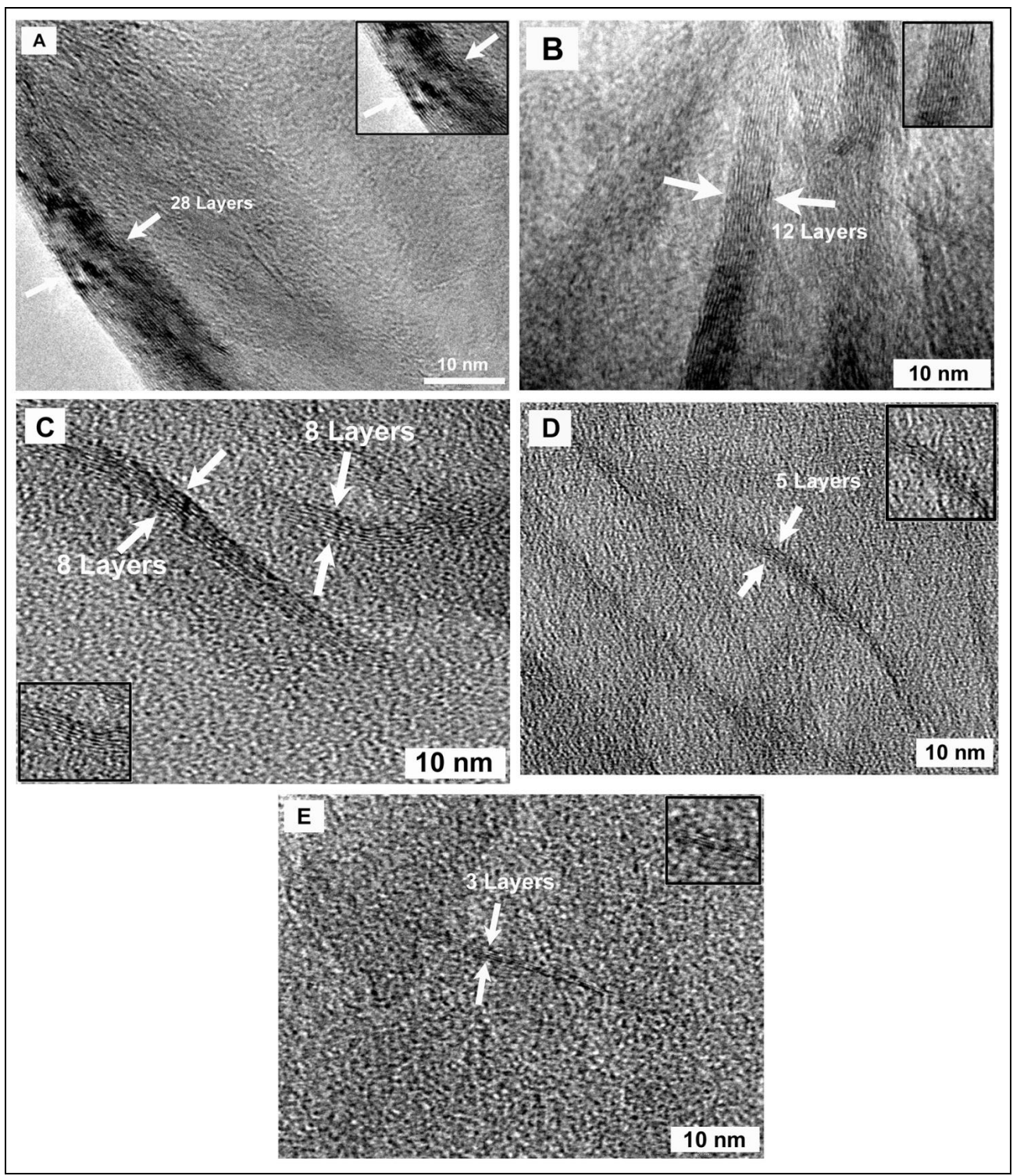

Fig. 3: HRTEM of the graphene stacked synthesized at (A) 80 atm Ar pressure, (B) 25 atm Ar pressure, (C) 1 atm Ar pressure (D) $10^{-3}$ torr and (E) $10^{-5}$ torr. The TEM images shows the corresponding numbers of sheets are 28, 12, 8, 5 and 3 respectively.

We will now describe a feasible mechanism based on the exfoliation of GO, under high (≥ 1 atm) external applied Ar gas pressure, to graphene stack of several graphene sheets and under low pressure (e.g $10^{-3}$ and $10^{-5}$ torr) to graphene stack of few graphene sheets, can be understood. The various carbon oxygen groups like epoxy (C-O-C), carbonyl (C=O), hydroxyl (O-H) and carboxylic (COOH) groups with sudden heating under Ar gas to ~1050 °C will break and combine to form $CO_2$. Since heating is sudden and the distribution of the various groups containing carbon and oxygen are not uniform, the $CO_2$ will be formed in local pockets. Because of the very low volume of local pockets ( ) where $CO_2$ gets formed. The pressure of $CO_2$ created inside the GO flake is very high. It is thought to be ~1100 atm (Schniepp et al. 2006). The $CO_2$ created in local pockets inside GO will create high isotropic pressure (~1100 atm). The pressure component parallel to the plane of the graphitic flake sheet will try to leave the stack. Also pressure component perpendicular to sheet will result in swelling and eventual breakage of the material. Now when there is external argon isotropic pressure of ≥ 1 atmosphere, the argon gas pressure from outside will counter the inside pressure. Therefore the exfoliation will lead to breakage of GO into elevated graphene stack.

In contrast to the above when externally applied Ar gas pressure is very small (e.g $10^{-3}$ and $10^{-5}$ torr), the high internal $CO_2$ pressure will lead to blast like situation in local regions. This will lead to a more effective exfoliation producing graphene stacks with lower number of graphene sheets. It can, therefore, be said that the number of graphene sheets in the exfoliated graphene stack will be large for high exfoliation pressure 80 atmosphere (28 sheets comparatively) small for moderate pressure e.g. 1 atmosphere (8 sheets) and few sheets for low pressure e.g 3 sheets for $10^{-5}$ torr pressure. Fig. 4 shows the schematic representation of pressure dependent thermal exfoliation of graphite oxide.

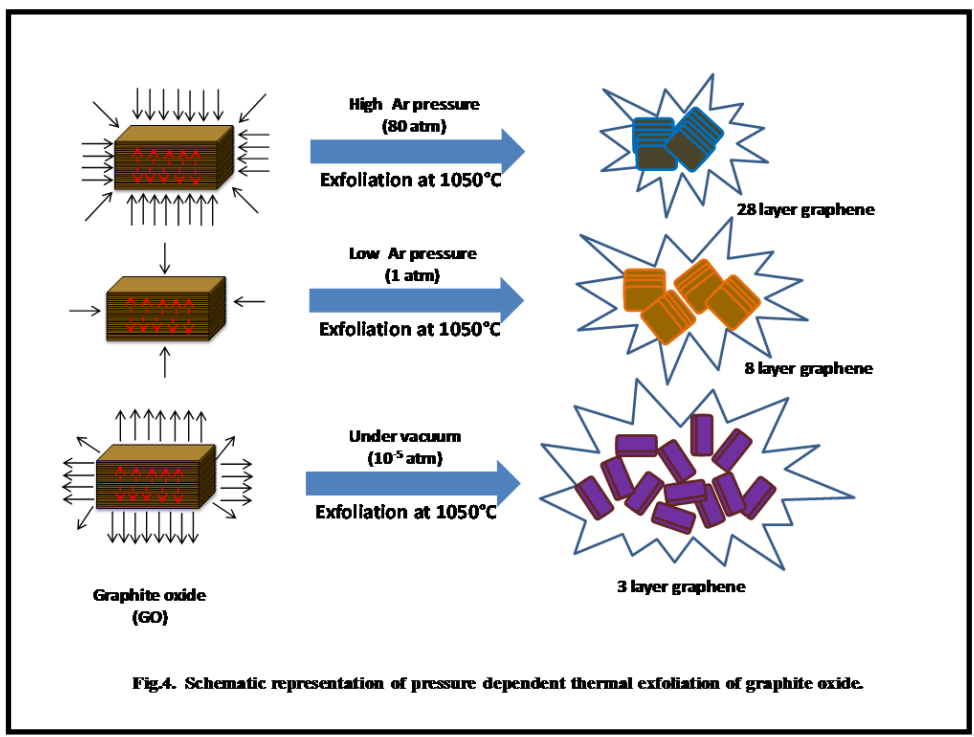

Fig.4: Schematic representation of pressure dependent thermal exfoliation of graphite oxide.

## 4. Conclusions

Effect of pressure on the thermal exfoliation of GO has studied. We have done thermal exfoliation of GO at high pressure (80 and 25 atm Ar), atmospheric pressure (1 atm Ar) and low pressure ($10^{-3}$ and $10^{-5}$ torr). The numbers of sheets in exfoliated graphene stack have been calculated by using Lorentzian fit of (002 peak) in XRD pattern. The number of sheets in graphene sheet has been found to be 28, 12, 8, 5 and 3 sheets at exfoliation pressures of 80, 25, 1 atmosphere and also at $10^{-3}$ and $10^{-5}$ torr pressure, respectively. Interestingly, we found that the formation of number of sheets in the exfoliated graphene stack is dependent on the applied exfoliation pressure on GO. By lowering the exfoliation pressure, the number of sheets gets reduced. This method provides the tuning of the number of graphene sheets for the exfoliation process.

## Acknowledgments

The authors are thankful to Prof. Lal Ji Singh (VC BHU), Prof. C.N.R. Rao and Dr. R. Chidambaram for their encouragement. The authors gratefully acknowledge to Prof. R.S. Tiwari, Dr. M.A. Shaz, Dr. T.P Yadav and Mr. D Pukazhselvan for their helpful discussions. Mr. Vijay Kumar and Mr. Vimal Kumar are acknowledged for their technical support in EM and XRD work, respectively. The author Himanshu Raghubanshi thanks the Ministry of New and Renewable Energy (MNRE) for providing the Senior Research Fellowship (SRF).